\documentclass[amssymb,amsfonts,aps,prb,twocolumn,longbibliography]{revtex4-2}

\usepackage{graphicx}
\usepackage{dcolumn}
\usepackage{bm}
\usepackage{bbold}
\usepackage[normalem]{ulem}
\usepackage{xcolor}
\usepackage{comment}
\usepackage{amsmath}
\usepackage{siunitx}
\usepackage{textcomp}
\usepackage{dsfont}
\usepackage{float}
\usepackage{braket}
\usepackage{placeins}
\usepackage[shortlabels]{enumitem}

\usepackage{appendix}

\allowdisplaybreaks

\begin{document}

\title{Exotic edge states of $C_3$ high-fold fermions in honeycomb lattices}

\author{
L. Madail$^{1,2}$,  R. G. Dias$^{2}$, J. Fern\'{a}ndez-Rossier$^{1}$\footnote{On permanent leave from Departamento de F\'isica Aplicada, Universidad de Alicante, Spain}
}
\affiliation{
$^1$ International Iberian Nanotechnology Laboratory (INL), Av. Mestre Jos\'{e} Veiga, 4715-330 Braga, Portugal
\\
$^2$ Department of Physics and I3N, University of Aveiro, 3810-193 Aveiro, Portugal
}

\date{\today}


\begin{abstract}
A generalization of the graphene honeycomb model to the case where each site in the honeycomb lattice contains a $n-$fold degenerate set of eigenstates of the $C_3$ symmetry has been recently proposed to describe several systems, including triangulene crystals and photonic lattices.  These generalized   honeycomb models are  defined by  $(n_a,n_b)$,  the number $C_3$ eigenstates in the $a$ and $b$ sites of  the unit cell, resulting in $n_a+n_b$ bands. Thus, the $(1,1)$ case gives the coventional honeycomb model that describes the two low-energy bands in graphene. 
Generalizations, such as $(2,1)$, $(2,2)$ and $(3,3)$ display 
several non-trivial features, such as coexisting graphene-like Dirac cones with flat-bands, both at zero and finite-energy, as well as robust degeneracy points where a flat-band and a parabolic band meet at the $\Gamma$-point. 
Here, we explore the edge states of this class of crystals, using as reference triangulene crystals,  and we find several types of edge states absent in the conventional $(1,1)$ honeycomb case, associated to the non-trivial features of the two-dimensional (2D) bands of the high-fold case.  First, we find dispersive edge states associated to the finite-energy flat-bands, that occur both at the armchair and zigzag termination. Second, in the case of non-centrosymmetric triangulene crystals that lead to a $S=1$ Dirac band, we have a bonding-antibonding pair of dispersive edge states, localized in the same edge so that their energy splitting is reduced as their localization increases, opposite to the  conventional behavior of pairs of states localized in opposite edges. Third, for the  $(3,3)$ case, that hosts a gap separating a pair of flat conduction and valence bands, we find non-dispersive edge states with $E=0$ in all edge terminations.

\end{abstract}

\maketitle

\section{Introduction}
It has long been known that the topology of the Fermi surface of materials  has a gigantic impact in their  electronic properties. The discovery of graphene promoted a way to rationalize our understanding of this interplay by analogy with massless Dirac fermions  \cite{novoselov05}. This made it possible to understand a number of graphene's exotic properties such as electron-hole symmetry  \cite{Yarmohammadi2016},  the unconventional Landau Level spectrum \cite{novoselov05},
Klein tunneling \cite{katsnelson06,beenakker08}, the finite value of its minimal conductivity \cite{katsnelson06b}, the universal value of the optical transmission coefficient \cite{nair08}, and the presence of zero-energy edge states in zigzag boundaries \cite{brey2006}.   

This paradigm has been extended more recently to the case of three-dimensional systems with diabolic points giving rise to Dirac cones, whose low-energy quasiparticles are isomorphic to Weyl fermions \cite{xu2015}. The common theme is the presence of degeneracy points in the band structure, that endows them with finite Berry curvatures and non-trivial quantum geometry  \cite{Rhim2020,Oh2022}. Unsurprisingly, analogy with a pre-existing relativistic theory is not necessary, and concepts such as {\em new fermions} \cite{bradlyn16} and {\em high-fold fermions} \cite{hasan21} has been put forward in order to dub exotic quasiparticles with non-trivial properties and without a relativistic analog. 

It was recently proposed by one of us \cite{ortiz22} a generalization of the graphene paradigmatic model, where quantum particles move in a honeycomb lattice, to the case where every site in the lattice hosts multiple orbitals that are eigenstates of the $C_3$ symmetry operator. The corresponding eigenvalues are given by $e^{i\nu \frac{2\pi}{3}}$,where $\nu=0,\pm 1$.
Here we shall refer to the quasiparticles in these systems as $C_3$ high-fold fermions, although the analysis will be relevant to bosonic quasiparticles as well.
The $C_3$ high-fold model provides an accurate description of the low-energy bands of two similar systems, 
the junctions of a crystalline network of graphene nanoribbons  \cite{Tamaki2020} and triangulene two-dimensional crystals \cite{ortiz22}, a honeycomb lattice whose unit cell contains a pair of $[n]$-triangulenes --  equilateral triangle shaped graphene nanoislands with a lateral dimension of $n$ benzene rings. Importantly,  $[n]$-triangulenes have a sublattice imbalance of  $N_A-N_B=n-1$, leading to the existence of $n-1$  sub-lattice polarized zero modes \cite{fernandez07}, that govern their low-energy properties and can be chosen as eigenstates of the $C_3$ symmetry operator \cite{ortiz19}.

Using both density-functional based calculations and the standard tight-binding (TB) model for graphene, it was found \cite{ortiz22} that honeycomb lattices, with a unit cell made of two triangulenes   dimensions  $n_a,n_b$, denoted in the following $[n_a,n_b]$-triangulenes,  have a set of $n_a+n_b-2$ low-energy bands, providing a realization of generalized honeycomb model with $(n_a-1,n_b-1)$. 
Hence, we found  that the low-bands of  $[n_a,n_b]$-triangulene crystals are described with the $(n_a-1,n_b-1)$ generalized honeycomb lattice, with the constraint that $\sum_{i=1,n} \nu_i =0$ for every triangulene.
Specifically, the calculations show that  \cite{ortiz22}:
\begin{enumerate}
    \item The two low-energy bands of the $[2,2]$-triangulene crystal were isomorphic to graphene, with reduced bandwidth, in agreement with previous work \cite{Tran2017}.
    \item The three low-energy bands of the $[2,3]$-triangulene crystal are isomorphic to the so-called $S=1$ Dirac model \cite{Mizoguchi2021} and the Lieb lattice \cite{Weeks2010d,Tsai2015}, with a single Dirac cone at the $\Gamma$-point, crossed at $E=0$ by a flat-band.
    \item The four low-energy bands of the $[3,3]$-triangulene crystal are isomorphic to the $p_x-p_y$ honeycomb lattice \cite{Congjun2007, Congjun2008}.
    \item The six low-energy bands of the $[4,4]$ are made of two pairs of Kagome-like bands, separated by a gap, where both the conduction and valence band are flat, in agreement with work by other authors
     \cite{Fang2024, Delgado2023}.
\end{enumerate}

Most of theory work in triangulenes has focused on their interacting properties associated to electron-electron interactions, for individual molecules \cite{fernandez07}, dimers  \cite{Mishra2020,Jacob2021,Henriques2023}, monostrand chains \cite{Mishra2021,Catarina2022,Martinez2023,Saleem2024} and  two-dimensional crystals  \cite{Zhou2020,Sethi2021,Catarina2023}. Very recently, the edge states of non-interacting electrons in the $[4,4]$-triangulene crystal have been also studied \cite{Fang2024}.
In this work we undertake the systematic study of edge states of the four representative classes of crystals listed above and relate their properties to the underlying isospin of the fermions in the low-energy bands of the two-dimensional lattice. For that matter, we adopt the same strategy of the seminal paper \cite{nakada96} on graphene edge states of Nakada {\em et al.} and we study the band structure of one-dimensional triangulene ribbons.  We do this at two different levels. First, we use the same tight-binding model of the two-dimensional lattice \cite{ortiz22}, for the complete atomistic model [see Fig.~\ref{fig:2}(a)].   Second, we use a reduced tight-binding model, that includes only the low-energy states of the triangulenes and maps them into an effective lattice.  This comparison permits us to ensure that the edge states found in the complete calculation are actually arising from the low-energy bands associated to the zero-modes of the triangulenes.

We stress here that our results will be of interest not only for the case of graphene, but also to any other system that can be described with a honeycomb lattice where every site hosts several modes with $C_3$ symmetry. Thus, very much like there are many {\em artificial graphene} systems, where honeycomb lattices host one orbital per site, realized with adatoms \cite{Sierda2023}, photons \cite{Plotnik2013}, polaritons \cite{Jacqmin2014}, surface electrons \cite{Gomes2012},  electric circuits \cite{Chen2023}, our work provides the generalization to analogous realizations of the multi-orbital $C_3$ symmetric honeycomb lattice. 

The rest of this work is organized as follows. In section II we introduce the complete tight-binding model that describes triangulene crystals as well as the reduced low-energy tight-binding models relevant for the $[2,2]$, $[2,3]$, $[3,3]$ and $[4,4]$ crystals. In addition, we review the evanescent Bloch states method to determine the dispersion of edge states and establish the topological invariant used to classify them. 
In section III, IV, V and VI we present the electronic structure for the four types of crystals, comparing the results for the complete and the reduced models, focusing on the properties of the low-energy states and on their topological regime.  
In section VII we summarize our main results.

\section{Methods}

    \subsection{Complete tight-binding model}

\begin{figure}
    \centering
    \includegraphics[width=1\linewidth]{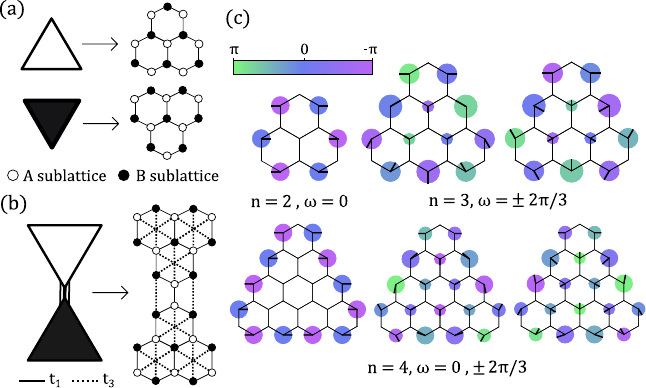}
    \caption{Single particle properties of graphene triangular nanoislands. (a) Example for $[2]$-triangulenes composed of (top) excess $A$ or (bottom) excess $B$  sublattice  sites. 
    (b) $[n_a,n_b]$-triangulene dimer.
    (c) Representation of the zero modes of $n=2,3,4$ triangulenes, chosen as eigenstates of $R_{2\pi/3}$ rotation operator with eigenvalues $e^{i\omega}$ and $\omega=0,\pm \frac{2\pi}{3}$.
    Circle size scales with the amplitude of the eigenfunction while the color represent their phase.}
    \label{fig:Fig1}
\end{figure}

The single particle properties of $[n_a,n_b]$-triangulenes, with one orbital per carbon atom on an unit cell dimer formed by two triangulenes of A and B types, is described by the following bipartite tight-binding Hamiltonian  
\begin{equation}
    H_{\text{TB}}=-t_1\sum_{<i,j>}a^\dagger_i b_j-t_3\sum_{\lll i,j\ggg}a^\dagger_i b_j,
    \label{eq:1}
\end{equation}
where $a^\dagger_i$ ($b^\dagger_i$) is the fermionic operator that creates an electron at site $i$ belonging to the $A$($B$)-sublattice. The first (second) sum is taken between first (third) neighbor sites in each triangulene whose vicinity is defined in Fig.~\ref{fig:Fig1} and includes all intercell hopping elements at the binding sites of the dimer. 
Following previous work \cite{ortiz22}, we assume $t_2=0$, so that the model is bipartite ensuring thereby  local charge neutrality, in agreement with density functional based calculations. 

The single-particle spectrum of individual $[n]$-triangulenes  described with equation (\ref{eq:1}) has a set of electron-hole symmetric states with finite-energy, well separated by a gap, and $n-1$ in-gap sublattice-polarized zero modes  with $E=0$. These modes can be chosen to be eigenvectors of the $C_3$ symmetry operator, on account of the point symmetry of triangulenes. 
Intermolecular hybridization among the zero modes of adjacent triangulenes, found for instance in density functional theory calculations, can only be established by third-neighbor hopping \cite{ortiz22}. This stems from the fact that the zero modes are hosted in the majority sublattice and the intermolecular binding sites belong to the minority sublattice. Thus, both $t_1$ and $t_2$ are linked to the binding sites where zero modes have zero probability amplitude and the leading term that produces intermolecular hybridization is $t_3$. 
    
It was found \cite{ortiz22} that a good agreement between spin-unpolarized DFT and tight-binding energy bands of $[n_a,n_b]$-triangulene 2D crystals is obtained if we take $t_3=0.1t_1$. These calculations display a wealth of weakly dispersive half-filled narrow bands located around the Fermi energy, well separated from higher/lower energy bands. Centrosymmetric $[n,n]$-triangulene crystals have $2(n-1)$ narrow bands. For $n\geq3$ they always feature pairs of electron–hole conjugate low-energy flat-bands, corresponding to states localized in supramolecular hexagonal rings. These flat-bands are very different from $E = 0$ bands, as they are not sublattice polarized and feature intermolecular hybridization. Non-centrosymmetric $[n_a,n_b]$-triangulene crystals feature $|n_a-n_b|$ narrow bands at $E=0$ due to the bipartite character of the lattice. More specifically, one is able to emulate a graphene-like spectrum for $n_a = n_b = 2$, spin-1 Dirac electrons for $n_a = 2,n_b = 3$, $p_{x,y}$-orbital honeycomb physics for $n_a = n_b = 3$, and a gapped system with flat valence and conduction bands for $n_a = n_b = 4$. \cite{ortiz22}.

    \subsection{Reduced model}
	
The total number of energy bands in the single-particle energy spectrum of $[n_a,n_b]$-triangulene ribbons is determined by the number of atoms  $N_r=\left[(n_a+2)^2+(n_b+2)^2-6\right]N_{rt}$ with $N_{rt}$ the number of triangulene dimers in the ribbon unit cell. For the 2D case, the $n_a+n_b-2$ weakly dispersive bands around $E=0$ are embedded inside a large gap isolated from higher energy bands. Therefore, low-energy physics at half-filling is expected to be governed by electrons occupying zero modes and these include all $n_{a/b}-1$ zero modes from each $[n_{a/b}]$-triangulene since intermolecular coupling is dominated by the small third-neighbor term. The effective 2D Bloch Hamiltonian in the basis of the $n_a+n_b-2$ zero modes as in  \cite{ortiz22} reads
\begin{equation}
	H_{\text{eff}}(\Vec{k})=
	\begin{pmatrix}
		0_{n_a-1} & \tau(\vec{k})\\
		\tau^\dagger(\vec{k})& 0_{n_b-1}
	\end{pmatrix}.
 \label{H2Def}
\end{equation}
This bipartite Hamiltonian describes sublattice polarized modes, $n_a-1$ ($n_b-1$) of those populate sites belonging to the $A$ ($B$) sublattice [see Fig.~\ref{fig:Fig1}(b)]. 
The elements of the $(n_a-1)\times(n_b-1)$ off-diagonal matrix $\tau(\vec{k})$ are given by
\begin{equation}
    \tau_{ab}(\vec{k})=\sum_{m=0,1,2}\tau^{ab}_{m}, \qquad \tau^{ab}_{m}=e^{i\vec{k}.\vec{R}_m}t_{m}^{ab},
    \label{eq:3}
\end{equation}
determined using the primitive vectors $\vec{R}_{1,2}=(n_a+n_b+1) \vec{a}_{1,2}$ that define the hexagonal lattice of the triangulene dimer in terms of the graphene primitive vectors $\Vec{a}_{1,2}$. Intracell ($m=0$) matrix elements are determined by the overlap between the components of the zero mode wavefunction at the binding sites. Meanwhile, the collection of states in the same sublattice reflect a rotational degree of freedom associated with $C_3$ symmetry. Inevitably, intercell ($m=1,2$) contributions can be expressed in terms of the previous ones using the phases ($\omega_{a/b}$) of the zero modes upon a $R_{2\pi/3}$ rotation operation as
\begin{equation}
    t_{1,2}^{ab}=e^{\pm i(\omega_b-\omega_a)}t_{0}^{ab}.
    \label{eqt12}
\end{equation}
We thus see that intermolecular hoppings, represented in the basis of $C_3$ eigenstates, can acquire a  Peierls-like phase factor \cite{ortiz22}.

It was found that the $n_a+n_b-2$ in-gap  energy bands of the complete tight-binding model of Eq.~\ref{eq:1} are in excellent agreement with the bands computed with the reduced model of Eq.~\ref{H2Def}.  As we discuss now,  the same statement applies also for the case of triangulene ribbons.

\subsection{Ribbon structures}

In order to study the edge states we compute the electronic structure of ribbons along the two main crystallographic directions of the honeycomb lattice, that generate edges with zigzag and armchair geometry  \cite{nakada96}.

The Hamiltonian of the $[n_a,n_b]$-triangulene ribbon  with $N_{rt}$ dimers in the ribbon unit cell can be expressed as the following banded block-Toeplitz matrix
\begin{equation}
    H_r(k)=\begin{pmatrix}
        \tau_{r,0} & \tau_{r,1}(k) &  & & & \\
        \tau_{r,1}^\dagger(k)& \tau_{r,0} & \tau_{r,1}^*(k) &  & & \\
         & \tau_{r,1}^\top(k)& \tau_{r,0} & \tau_{r,1}(k) & \\
         &  & \tau_{r,1}^\dagger(k)& \tau_{r,0}&\ddots \\    
         & & & \ddots&\ddots\\
    \end{pmatrix}.
    \label{Hrk}
\end{equation}
where $\tau$ are matrices. Within the full tight-binding approach, the dimension of each $\tau_{r,\sigma}$ entry is $N_{t_a}\times N_{t_b}$ with $N_{t_{a/b}}$ the number of atoms in each $[n_{a/b}]$-triangulene, making the Hamiltonian a total of $(N_{t_a}+N_{t_b})N_{rt}\times(N_{t_a}+N_{t_b})N_{rt}$ matrix.
In contrast, within the reduced tight-binding approach, the dimension of the $\tau_{r,\sigma}$ matrices reduces to $(n_a-1)\times (n_b-1)$ if we consider the effective low-energy model on the basis of the zero modes. Block-diagonal $\sigma=0$ entries account for couplings within the ribbon unit cell and are block anti-diagonal following Eq.~\ref{H2Def} while off-diagonal $\sigma=1$ matrices connect adjacent unit cells. Their corresponding elements for a zigzag ribbon can be explicitly written in terms of the established entries of the 2D effective Bloch Hamiltonian as
\begin{equation}
   \begin{aligned}
    \tau_{r,0}^{ab}&=\tau_0^{ab},\\
    \tau_{r,1}^{ab}&=\tau_1^{ab}(k_x=0,k_y=0)+\tau_2^{ab}(k_x,k_y=0),
    \end{aligned}
\end{equation}
where the momentum $k_x$ has been renormalized to the width of each ribbon unit cell.

In both edge configurations, the atoms on the triangular islands that populate the boundaries belong to both sublattices. Since dimers connect two triangulenes with opposite majority sublattices, sublattice imbalance between edges is entirely suppressed in armchair ribbons as well as in zigzag ribbons provided that $n_a=n_b$. The top (bottom) zigzag edge is composed of $[n_a]$-triangulenes ($[n_b]$-triangulenes) with extra $(n_a-1)$ [$(n_b-1)$] atoms in the majority sublattice, while both armchair edges are made of triangulene dimers with extra $|n_a-n_b|$ atoms in the majority sublattice.

\subsection{Evanescent Bloch waves for the calculation of edge states}\label{sec:Bloch}

We now review a method  \cite{ Bellec2014, Kohmoto2007, Laude2009, Alase2017, Cobanera2018, Kunst2019h} to obtain the energy dispersion and the wave function of edge states, based on the generalization of the Bloch states to the case of imaginary wave vectors. Conventional Bloch states are eigenstates of the discrete translation operator that leaves the crystal invariant. In a 1D crystal, relevant for our case, we define the translation operator as
 $\mathcal{T}=\sum_{j\in\mathbf{Z}}\ket{j}\bra{j+1}$, where $j$ identifies the unit cells of the crystal. We label the eigenstates and eigelvalues as $\mathcal{T}\ket{z}=z\ket{z}$ with
\begin{equation}
    \ket{z}_\gamma=\sum_{j\in \mathbb{Z}} z^j\ket{j} , z\in\mathbb{C}.
\end{equation}
 Conventionally, the eigenvalues $z$  are phases $e^{i k }$, where $k$ is chosen as to match the periodic boundary conditions. In this case, repeated application of the translation operator results in a phase modulation of the wave function, without affecting the amplitude. Analytical extension to the complex plane permits one to define real eigenvalues, $z$, that result in evanescent/exploding solutions, depending on whether $z<1$ or $z>1$.
These non-unitary representations of the states can  only describe edge states.

Using the standard ansatz in $\langle j |\psi_z(j)\rangle= z^j |u(z)\rangle$, we need to solve 
\begin{equation}
    h(z)\ket{u(z)}=\epsilon(z)\ket{u(z)},
    \label{eqhz}
\end{equation}
where  $h(z)$ is the usual square matrix of dimension $n$,  the number of sites of the crystal cell.  The conventional Brillouin zone (BZ) is the unit circle with $|z|=1$ in the complex plane. 

For $|z|\neq1$, $h(z)$ is non-Hermitian and its eigenstates have an exponential behavior. The full set of solutions of the bulk equation is obtained by requiring that the corresponding eigenspectrum is real. The diagonalization of Eq.~\ref{eqhz} yields two roots $\{z_n\}$ for a given energy $\epsilon$. Since we are looking for solutions of the open system, the boundary equation is introduced at this stage to determine which $\{z_n\}$ belong to the energy spectrum of $H$. 

For a given $k_x$, the Hamiltonian of the ribbons defines a finite-size one-dimensional chain. We can therefore, implement this approach for every $k_x$ and find thereby the analytic form of the edge modes of triangulene ribbons assuming the generalized solutions only in the open boundary direction ($\gamma=x$) while using the Bloch theorem in the periodic direction to be valid in the bulk region, thus obtaining the energy dispersion $\epsilon(z_n,k)$. The allowed $\{z_n\}$ solutions at the edge are obtained by addressing the boundary equation, whose definition may not be straightforward in models with multiple orbitals per unit cell. In the following sections, we make this boundary equation self-evident by means of a basis rotation via bonding and anti-bonding linear combination of the Wannier states associated with the zero modes of each $[n]$-triangulene.


 \subsection{Topological invariant}

The notion of the Zak phase for one-dimensional models is well defined as the integral of the Berry connection across the Brillouin zone  \cite{Zak1989j}. In two-dimensional systems, this integration should be taken on a cut of a 2D Brillouin zone in a direction transverse to the ribbon orientation. Translation invariance along the direction $k_\parallel$ parallel to the boundary guarantees a continuous parameter in space in which the state $\ket{u(\vec{k})}$ can travel on a closed path. The parallel component is real and still a good quantum number, while the normal component is complex corresponding to the exponential decay factor  \cite{Delplace2011}.  Therefore, we can define the Zak phase $\mathcal{Z}(k_\parallel)$ of an isolated band as the integration of the Berry connection across the 2D Brillouin zone along the perpendicular direction $k_\bot$ in
\begin{equation}
    \mathcal{Z}(k_\parallel)=i\oint \braket{u(\vec{k})|\partial_{k_\bot}u(\vec{k})} dk_\bot.
    \label{zak1}
\end{equation}
For finite but sufficiently long periodic systems ($N_{rt}\gg 1$), the numerical computation of Eq.~\ref{zak1} can be carried out using a discrete set of $k_\bot$-points in the first BZ as follows
\begin{equation}
    \mathcal{Z}(k_\parallel)=-\operatorname{Im}\operatorname{ln}\prod_{j}^{N_{k_\bot}-1}\braket{u(k_{j})|u(k_{j+1})}.
\end{equation}
Chiral and inversion symmetric models have quantized Zak phases that can only take a trivial or non-trivial value ($0\,\wedge\,\pi\mod2\pi$).  According to the bulk-boundary correspondence, there are low-energy excitations at the interface between topologically distinct insulators with a trivial and a non-trivial phase. This means that, when the phase is $\pi$-quantized, a ribbon with open boundary conditions along $k_\bot$ (interface with vacuum) and invariant by translation along $k_\parallel$ supports interface modes.

In the multiband formalism, the topological property of a bandgap can be determined by the summation of the $\mathcal{Z}_n(k_\parallel)$ phases of the set of bands below this gap  \cite{Xiao2015,Wang2016}. Specifically for insulating inversion symmetric models and as $k_\bot$ is integrated along the path perpendicular to the edge, the total phase $\sum_{n\in \text{occ}}\mathcal{Z}_n(k_\parallel)/\pi$ of the occupied bands is equivalent to the winding of the trajectory of $\det \tau (k_\parallel)$ and predicts the number of pairs of flat edge states in the midgap  \cite{Wang2016, Milicevic2017}. Here, we use this formalism to predict the topological phase of all bandgaps for each triagulene ribbon, including the non-central ones, where pairs of finite-energy dispersive edge states are found. The case of degenerate Hamiltonians can be treated using the non-Abelian formalism which measures the geometric phase associated with the matrix of a subspace of the Hilbert space  \cite{Sarandy2006, Sugawa2021}. However, this approach is not able to predict edge states between degenerate bands. Herein, the presence of singular points of degeneracy for each $k_\parallel$ was treated by making use of a smooth gauge choice where phases are roughly uniformly distributed throughout the loop. In the case where  there is a finite number of degeneracies in the band structure for each $k_\parallel$, this gauge allows for these singular points to be excluded from the numerical calculation of $\mathcal{Z}_n(k_\parallel)$ without significantly affecting the results. Additional considerations on this approach can be found in appendix \ref{SMB}.

	\section{$[2,2]$-triangulene ribbons}

\begin{figure}[b]
     \centering
     \includegraphics[width=1\linewidth]{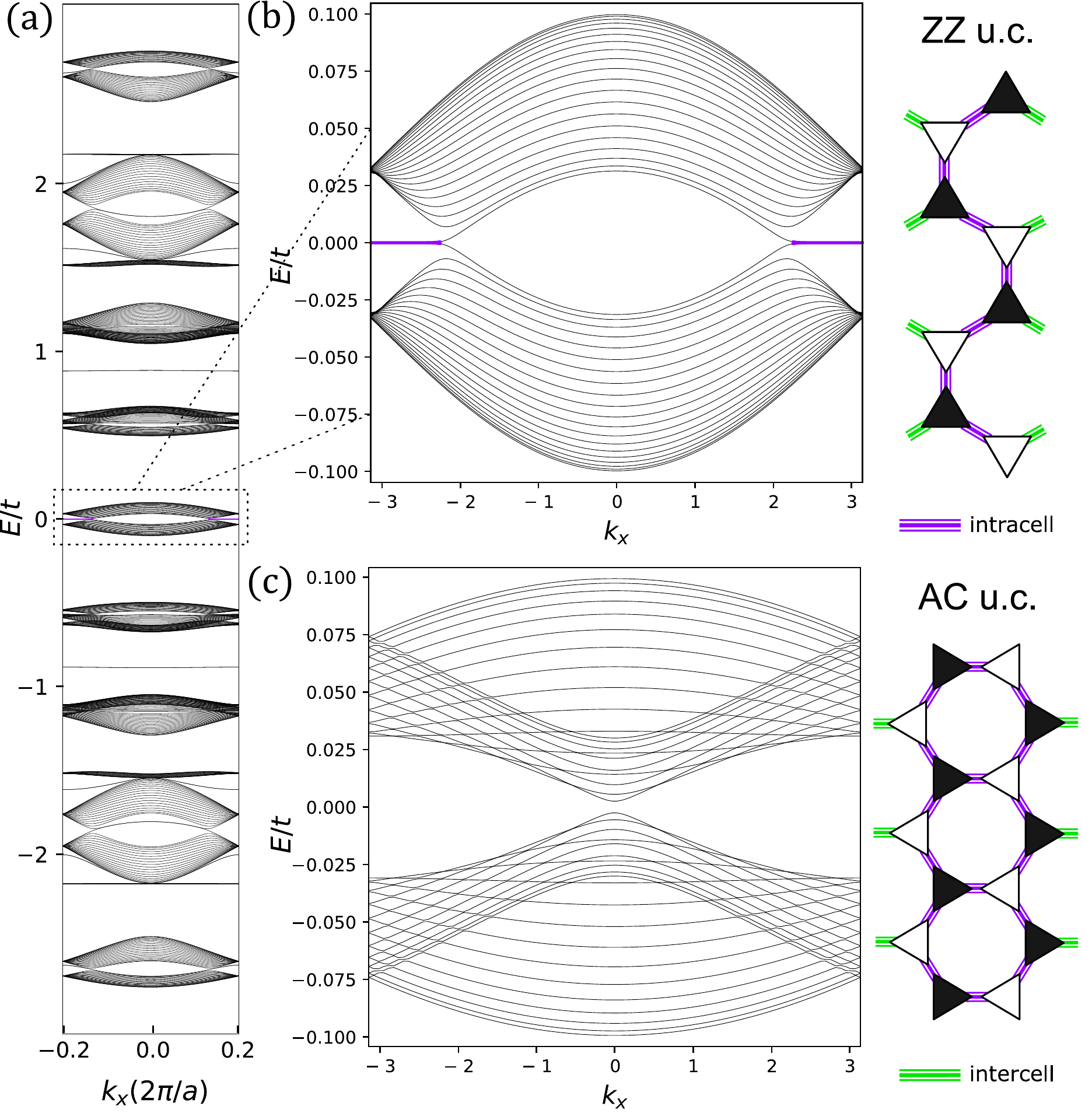}
     \caption{Full model for $[2,2]$-triangulene ribbons in the zigzag and armchair configurations. The full tight-binding diagonalization reveals a total of $N_r=26N_{rt}$ bands, as shown in the spectrum (a) for a $[2,2]$-triangulene zigzag ribbon with $N_{rt}=20$ dimers, $2N_{rt}$ of those isolated around zero-energy. A closeup near this point signals an identical spectrum to the reported graphene ribbons, both in the (b) zigzag and (c) armchair configurations.}
     \label{fig:2}
 \end{figure}

The results for $[2,2]$-triangulene ribbons in both the zigzag and armchair configurations with $N_{rt}=20$ triangulene dimers, with 26 sites per dimer,  in the unit cell are shown in  Fig.~\ref{fig:2}. The full tight-binding models yield $N_r=26N_{rt}$ bands, wherein $(n_a+n_b-2)N_{rt}=2N_{rt}$ of those are located around the Fermi energy, well separated from the remainder $24N_{rt}$ higher/lower energy bands [see Fig.~\ref{fig:2}(b)]. 

The configuration for triangulene ribbons is established by replacing the unit cell dimer of the triangular Bravais lattice with a triangulene dimer [see Fig.~\ref{fig:2}(a)]. In fact, every $[2]$-triangulene hosts exactly one isolated zero mode and intermolecular coupling between these states is solely assured by $t_3$. Thus, one expects the low-energy bands for $[2,2]$-triangulene ribbons to be isomorphic to the tight-binding model for a honeycomb ribbon with nearest neighbor approximation and one orbital per site, provided that we keep the analogy in the choice of open boundary conditions  \cite{Lado2015}. Nevertheless, in constrast with the models of graphene with nearest-neighbor approximation, the narrower low-energy bandwidth of triangulene ribbons is given by the smaller amplitude of the third-neighbor hopping term. Therefore, full concurrence between the low-energy bands of $[2,2]$-triangulene ribbons and the $p_z$-orbitals of graphene ribbons can be obtained for an effective hopping term of $\tilde{t}_g=\frac{t_3}{3}$ with energy bandwidth of $6\tilde{t}_g$. 

In the zigzag configuration, the boundary condition forces the wave functions to vanish in a single sublattice at each edge such that sublattice symmetry is broken and the two Dirac cones project onto inequivalent points $k_x=\pm 2\pi/3$ on the edge. This gives rise to two zero-energy edge states when $|k_x|\in [2\pi/3,\pi]$. 
These states are localized at the edges, analog to the extensively studied zigzag edge states in graphene \cite{nakada96,Lado2015}.  
The single band formalism was employed here to calculate the Zak phase of the band below the midgap. We find a topological non-trivial phase $\mathcal{Z}=\pi$ for the interval $|k|\in [2\pi/3,\pi]$ and a trivial phase otherwise, effectively predicting the pair of zero-energy edge states.  In the armchair configuration, we also obtain a set of low-energy bands isomorphic to that of armchair graphene ribbons  [see Fig.~\ref{fig:2}(c,d)]. 

The analogy  between the zigzag edge states of the $[2,2]$-triangulene crystals and those of graphene  is not perfect when we analyse the wave functions. In the case of graphene, the amplitude of the edge mode associated to the zigzag edge with $A$ atoms vanishes in the $B$ sublattice.  In contrast, in the triangulene case, the wave function of the edge states of the $A$-rich side has a small weight in the $B$ rich triangulenes. This departure is only captured in the complete model, but not in the reduced model. Therefore, it has to be originated by  a small influence of the  higher/lower energy bands.

\section{$[2,3]$-triangulene ribbons}

In this section, we present the energy dispersion of $[2,3]$-triangulene ribbons using the complete tight-binding model and compare with the reduced effective model that only includes the  zero modes. We also analyse  the localization and topological properties of edge modes found specifically in zigzag ribbons, while considerations on armchair edge states are provided in appendix \ref{SMA}.  

 \subsection{Complete model}

In Fig.~\ref{fig:3} we show the low-energy bands corresponding to zigzag and armchair ribbons, calculated with the complete tight-binding model. These low-energy bands are identical to those obtained for the same geometries using the minimal model, confirming that they are linear combinations of the zero modes of the triangulenes.

We find three types of states for the $[2,3]$ ribbons.
First, zero-energy states associated to the sublattice imbalance of the $[2,3]$ dimer. Specifically, we find as many $E=0$ states as $[2,3]$ dimers in the unit cell. Therefore, only two of these states are localized at the edges. 
Second, an electron-hole symmetric pair of dispersive edge states [see purple lines in Fig.~\ref{fig:3}(a,b)], different from the graphene case in several counts:
\begin{enumerate}
    \item They are found both for zizag and armchair terminations in contrast with graphene, where armchair boudaries do not have edge states;
    \item For the zigzag direction, these states are only localized in the edge terminated with the $[2]$-triangulene [see Fig.~\ref{fig:3}a(i)];
    \item For the armchair configuration,  edge states [colored bands in Fig.~\ref{fig:3}(b)] are found in both sides of the ribbon and there are two pairs of electron-hole symmetric edge states specifically at the zone boundary;
    \item The localization of the all edge states is maximal at the zone boundary, which leads to a maximal electron-hole splitting, as a result of the inter-edge nature of the states;
    \item The dispersive edge bands never reach $E=0$.
    In the next subsection we derive an analytical formula for the dispersion of the zigzag edges states by applying the evanescent Bloch wave method to the reduced tight-binding model;  
    \item They cannot be predicted by the Zak phase, that is no longer quantized  in the absence of inversion symmetry,  a characteristic of the $[2,3]$ non-centrosymmetric triangulenes.
\end{enumerate}

The third type of state that we found is the group of dispersive bulk modes associated to the quantum confined subbands of the Dirac cone. Importantly, because of the confinement induced by the finite width of the ribbons, there is a gap at the $\Gamma$-point, in contrast with the 2D crystal. In the neighborhood of $\Gamma$ we have found that the energy of the confined states can be described by

\begin{equation}
    \epsilon(k_x,n)=\pm \hbar v_F \sqrt{k_x^2 + \left(\frac{\pi\left(n-\frac{1}{2}\right)}{W\left(N_{rt}+1\right)}\right)^2},
    \label{e23_1}
\end{equation}
with $W$ the width of the ribbon and the Fermi velocity $\hbar v_F^{[2,3]}= 6\sqrt{\frac{3}{11}}t_3a_{cc}$ already established in the literature \cite{ortiz22}.  In the next subsection we derive this formula, making use of the reduced tight-binding model.

 \subsection{Reduced model}

 \begin{figure*}[t]
     \centering
     \includegraphics[width=1\textwidth]{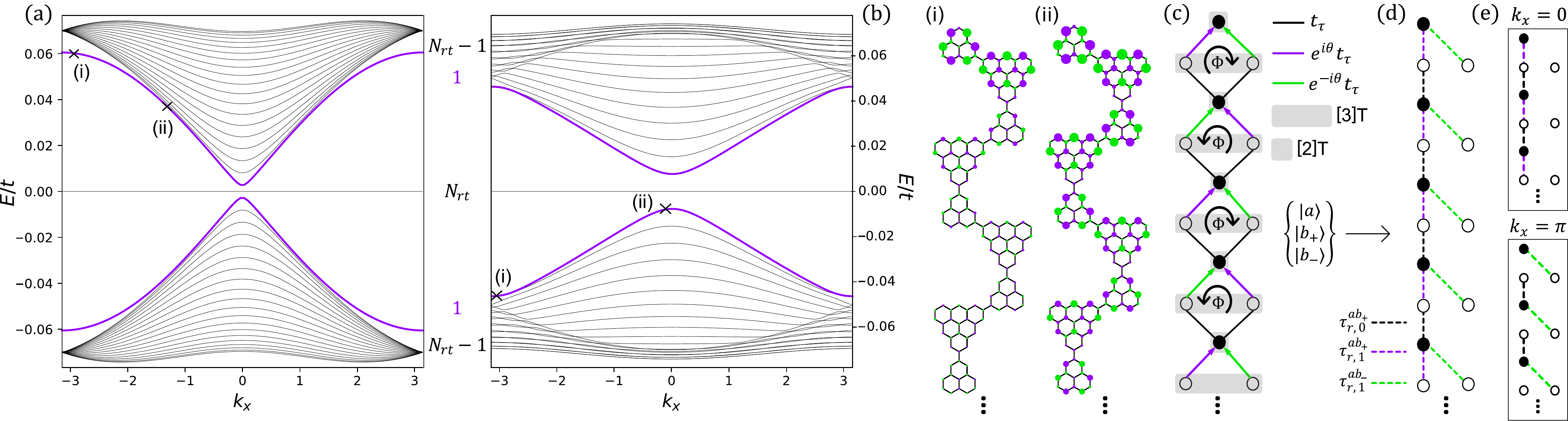}
     \caption{Modelling $[2,3]$-triangulene ribbons with $N_{rt}=20$ dimers in the unit cell. The numerical results of the energy dispersion for the (a) zigzag and (b) armchair configurations show the presence of midgap edge states highlighted in purple. In the former, the states localize in (i) the top edge and (ii) decay through the bulk away from the zone boundary. The zigzag ribbon unit cell effectively maps into a (c) diamond chain in the $C_3$ symmetric basis and a (d) SSH model with dangling sites in the bonding/anti-bonding combination basis, whose unit cells at the high-symmetry points are depicted in (e). }
     \label{fig:3}
 \end{figure*}
 
We now shed light on some of the results of Fig.~\ref{fig:3}(a)  using the reduced tight-binding model. In the case of $[2,3]$-triangulenes, the minimal model considers $n_a+n_b-2=3$ zero modes per triangulene dimer, one in the $[2]$-triangulene, denoted as $\ket{a}$, two in the $[3]$-triangulene, denoted as $\{\ket{b_1},\ket{b_2}\}$. Within the reduced tight-binding model, the Hamiltonian matrix of a $[2,3]$ zigzag ribbon made of $N_r$ dimers, has dimension $3*N_r$. The entries of this matrix are given by the intra- and inter-cell contributions (see Eq.~\ref{eq:3})
\begin{equation}
    \tau_{r,0}=\begin{pmatrix}
    0& \tau_{r,0}^{ab_1}& \tau_{r,0}^{ab_2}\\
    \tau_{r,0}^{ab_1\dagger}& 0 & 0\\
    \tau_{r,0}^{ab_2\dagger} & 0 & 0
    \end{pmatrix}, \;\;
    \tau_{r,1}=\begin{pmatrix}
    0& 0& 0\\
    \tau_{r,1}^{ab_1}& 0 & 0\\
    \tau_{r,1}^{ab_2} & 0 & 0
    \end{pmatrix},
\end{equation}
whose entries follow
\begin{equation}
    \tau_{r,0}^{ab_{1,2}}=t_{[2,3]},\quad 
    \tau_{r,1}^{ab_{1,2}}=\left(e^{i\theta}+e^{\mp i\theta}e^{-ik_x}\right)t_{[2,3]},
\end{equation}
with $t_{[2,3]}=\frac{2t_3}{\sqrt{6\times11}}$ and $\theta=\frac{2\pi}{3}$. The structure of the $\tau_{r,\sigma}$ matrices permits one
to map the low energy states of the $[2,3]$ triangulene zigzag ribbon unit into a finite-size diamond chain, where every site in the diamond lattice represents a zero mode state. Within this effective model, hoppings have additional $e^{i\theta}$ Peierls-like factors, amounting to a magnetic flux within the unit cell of $\phi_{uc}=\frac{\phi_0}{2\pi}2\theta=\frac{2}{3}\phi_0$ per plaquette, normalized to $\phi_0=h/e$ [see Fig.~\ref{fig:3}(c)]. Naturally, the effective magnetic flux within a plaquette is $k$-dependent due to the $e^{\pm ik_x}$ phase factors in $\tau_{r,1}$ matrices.  We now perform a basis rotation using bonding/anti-bonding linear combination between the two zero modes localized in the B-type $[3]$-triangulene. This breaks $C_3$ symmetry between the eigenfunctions.  In the new basis set $\{\ket{a},\ket{b_+},\ket{b_-}\}$, the $\tau_{r,\sigma}^{ab^\pm}$ matrices now satisfy
\begin{equation}
   \begin{aligned}
    \tau_{r,0}^{ab^\pm}&=(1\pm1)t_{[2,3]},\quad  \tau_{r,1}^{ab^+}=\left(-1-e^{-ik_x}\right)t_{[2,3]},\\
    \tau_{r,1}^{ab^-}&=\left(-1+e^{-ik_x}\right)\sqrt{3}it_{[2,3]}.
    \end{aligned}
\end{equation}

As one of the intra-cell hopping terms vanishes, the antisymmetric combination of states $\ket{b_-}$ become coupled only to adjacent triangulene dimers [see Fig.~\ref{fig:3}(d)]. These states get fully decoupled from the chain at the $\Gamma$-point where $\tau_{r,1}^{ab^-}=0$ and contribute with $N_{rt}$ zero-energy bands. At the $k_x=0$ point, the system behaves as an open linear chain that couples the $2N_{rt}$ bonding states $\ket{a(b)_+}$ through an uniform hopping term $\tau_{r,1}^{ab^+}=\tau_{r,0}^{ab^+}=2t_{[2,3]}$ [see Fig.~\ref{fig:3}(e)]. Note that we have dropped the relative phase between $\tau_{r,0}^{ab^+}$ and $\tau_{r,1}^{ab^+}$ since the closed loops disappear for $k_x=0$. Thus, the energy of the dispersive subbands is given by
\begin{equation}
    \epsilon(k_x=0,k_n)=4t_{[2,3]} \cos(k_n),
    \label{eq:15}
\end{equation}
where $k_n=\frac{\pi}{2N_{rt}+1}n$ and $n=1,\dots,2N_{rt}$. It can be seen that Eq.~\ref{eq:15} is consistent with Eq.~\ref{e23_1} evaluated at $k_x=0$.

Analytical expressions can also be obtained at the zone boundary $(k_x=\pi)$. Specifically, the three highly degenerate points with energies $\epsilon(k_x=\pi)=\{0,\pm 4t_{[2,3]}\}$ are due to the trimerization depicted in Fig.~\ref{fig:3}(e), on account of the vanishing $\tau_{r,1}^{ab^+}$ entries. 
As a result, their eigenfunctions have  zero amplitude in the $[2]$-triangulene of each trimer. Additionally, at the open boundary filled with $[2]$-triangulenes, a dimer is formed leading to the chiral pair with energies $\epsilon_\text{edge}(k_x=\pi)=\pm 2\sqrt{3}t_{[2,3]}$. In the opposite boundary, a single decoupled site located in the $[3]$-triangulene contributes with a zero-energy edge state. 

The analytical form of the finite-energy edge state can be recovered by assuming a generalized solution in the open $y$-direction, as introduced in section \ref{sec:Bloch}. We look for suitable solutions of the form $\ket{\epsilon_\text{edge}}=e^{ik_x}z_y^j\ket{u(z)}$ in the basis set of the zero modes $\{\ket{a},\ket{b_1},\ket{b_2}\}$ for the following bulk Hamiltonian $h(k_x,z_y)$ of a $[2,3]$-triangulene crystal
\begin{equation}
    h(k_x,z_y)=
    t_{[2,3]}\begin{pmatrix}
    0&\tau_{ab_1} &\tau_{ab_2} \\
    \tau_{ab_1}'& 0 & 0\\
    \tau_{ab_2}' & 0 & 0
    \end{pmatrix}, \label{h23edge23}
\end{equation}
with entries
\begin{equation}
   \begin{aligned}
\tau_{ab_{1,2}}&=1+2z_y\cos(R_{1_x}k_x \pm \theta),\\
\tau_{ab_{1,2}}'&=1+2z_y^{-1}\cos(R_{1_x}k_x \pm \theta),
    \end{aligned}
\end{equation}
and $R_{1_x}=(n_a+n_b+1)a/2$. The allowed $\{z_{y_n}\}$ solutions are obtained by requiring that the eigensolutions of the Hamiltonian in Eq.~\ref{h23edge23} are also admissible in the open system. Zigzag open boundary conditions require breaking in half the dimer unit cells at the edges. The corresponding boundary equations can be easily identified in the rotated mapping version of Fig.~\ref{fig:3}(d), where the number of intracell hopping terms is minimized. Given that the chiral pair only localizes in the $[2]$-triangulene-terminated  zigzag edge, the relevant boundary equation imposes that the wave function vanishes at $b_+$ sites outside the ribbon, that is $\braket{b_+|\epsilon_\text{edge}}=0$. This leads to the only admissible solution $z_y=\cos(R_{1_x}k_x)$, resulting in the following analytical form of the energy dispersion for this pair of edge states
\begin{eqnarray}
    \epsilon_\text{edge}=\pm \sqrt{6}t_{[2,3]}\sqrt{1-\cos^2(R_{1_x}k_x)}=\nonumber\\
    =\pm \sqrt{6}t_{[2,3]}\sqrt{1-z_y(k_x)^2},
    \label{eq:18}
\end{eqnarray}
that shows how the splitting of the edge states is maximal when the localization is maximal, $z_y=0$. This is at odds with conventional edge states in zizgzag graphene, on account of the fact that the two edge states of the $[2,3]$ zigzag ribbon live on {\em the same edge}. Therefore, their hybridization is reduced when their localization decreases, which happens as $k_x$ moves from the zone boundary towards the zone center.  Eq.~\ref{eq:18} is consistent with the numerical calculations except around the $\Gamma$-point, where the edge state becomes more extended and the effect of the other boundary condition becomes relevant.

\section{$[3,3]$-triangulene ribbons}

 We present the results for the energy dispersion $[3,3]$-triangulene ribbons using the tight-binding complete model and the effective reduced model. We focus our analysis on the localization and topological properties of edge modes found specifically in zigzag ribbons, while considerations on armchair edge states are provided in appendix \ref{SMA}. 
 
 \subsection{Complete model}
We now consider the edge states of $[3,3]$-ribbons. This parent 2D crystal has four sets of low-energy bands that arise from the four zero modes of the unit cell. Two of them are isomorphic to graphene bands, the other two are finite-energy flat-bands, associated to ring states \cite{Congjun2007,Tamaki2020,ortiz22}, degenerate with the dispersive bands at the $\Gamma$-point. Therefore, we should expect edge states associated to the graphene-like bands, as well as edge states associated to the flat-bands. In the ribbon geometry, we find a total of $4N_{rt}$ low-energy bands, where $N_{rt}$ is the number of $[3]$-triangulene dimers in the unit cell, organized in four groups of low-energy bands plotted in Fig.~\ref{fig:4}(a,b).

\begin{itemize}
    \item[-] The first group of flat bands with finite energy, that belong to ring states, i.e, to  confined states in the smallest closed loops made of six $[3]$-triangulenes in a hexagonal plaquette. The number of states of this type is 
    $2(N_{rt}-1)$ for zigzag ribbons and $2(N_{rt}-2)$  for  armchair ribbons. 
    \item[-]  The second group has $2(N_{rt}-2)$ electron-hole symmetric dispersive bands. Two Dirac cones at $k_x=\pm2\pi/3$ in zigzag ribbons and valley admixing in armchair ribbons are predicted in these and all centrosymetric triangulenes where 3-fold symmetry is preserved.   At the $\Gamma$-point, the bands of the second group become   becomes degenerate with the ring states, very much like in the 2D crystal.
    \item[-] A third group of two (four) edge-state bands for the zigzag (armchair) geometry, purple highlighted in Fig.~\ref{fig:4}(a) [\ref{fig:4}(b)] in the gap of the second group of dispersive bands. For the zigzag configuration, the pair of bands have zero-energy and are localized at the $[3]$-triangulene in both boundaries when $0<|k|<2\pi/3$ [see Fig.~\ref{fig:4}a(i)]. This interval is complementary to the one found in the edge states of $[2,2]$-triangulene ribbons and emulates the behavior of edge states in bearded zigzag graphene  \cite{Plotnik2013}. In armchair ribbons, this pair of dispersive states is also associated to edge states, maximally localized at the zone boundary in both sides of the ribbon, populating the outermost dimers (see appendix \ref{SMA}).
    \item[-] The fourth group, highlighted in green in Fig.~\ref{fig:4}(a) [Fig.~\ref{fig:4}(b)], consists of two degenerate pairs of states associated to the band touching points between the first and second group of bands. These are present in both boundaries and are compactly localized in one (two) edge dimers for $k_x=0$ ($k_x=\pi$). In the vicinity, these states acquire a $k$-dependent $c_y$ decaying factor before loosing its localization properties at the $k_x=\pi$ ($k_x=0$) point [see Fig.~\ref{fig:4}a(ii)].
\end{itemize}

\subsection{Reduced model}

\begin{figure*}[t]
     \centering
     \includegraphics[width=1\textwidth]{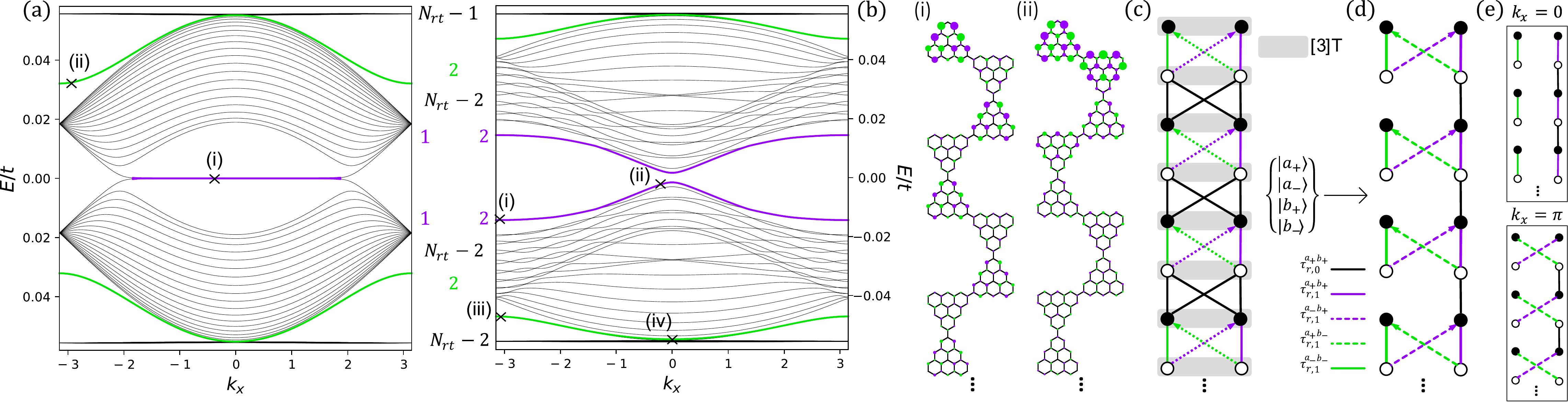}
     \caption{Model for $[3,3]$-triangulene ribbons with $N_{rt}=20$ dimers in the unit cell. The numerical results of the energy dispersion for the (a) zigzag and (b) armchair configurations show the presence of two types of midgap edge states. In the former, states highlighted in purple [green] localize in a (i) single triangulene [(ii) triangulene dimer] at the top and bottom boundaries. The zigzag ribbon unit cell effectively maps into a (c) Creutz ladder in the $C_3$ symmetric basis. The number of intracell hopping terms is reduced if one considers (d) the bonding/anti-bonding combination basis, whose unit cells at the high-symmetry points are depicted in (e).}
     \label{fig:4}
 \end{figure*}

Individual $[3,3]$-triangulene dimers host $4$ zero modes that form the basis set $\{\ket{a_1},\ket{a_2},\ket{b_1},\ket{b_2}\}$ for the internal space of the effective zigzag ribbon Hamiltonian. This set can be identified by their sublattice occupation and relative phase $\omega_{1,2}=\pm2\pi/3$ between eigenstates of the rotation operator $R_{2\pi/3}$ within the same sublattice. The finite elements of the effective zigzag ribbon Hamiltonian are given by the intra and intercell matrices following
\begin{equation}
   \begin{aligned}
    \tau_{r,0}&=\begin{pmatrix}
    0&0& \tau_{r,0}^{a_1b_1}& \tau_{r,0}^{a_1b_2}\\
    0&0& \tau_{r,0}^{a_2b_1}& \tau_{r,0}^{a_2b_2}\\
    \tau_{r,0}^{a_1b_1\dagger}& \tau_{r,0}^{a_2b_1\dagger} & 0&0 \\
    \tau_{r,0}^{a_1b_2\dagger} & \tau_{r,0}^{a_2b_2\dagger} & 0&0 
    \end{pmatrix}, \\
    \tau_{r,1}&=\begin{pmatrix}
    0&0& 0& 0\\ 
    0&0& 0& 0\\
    \tau_{r,1}^{a_1b_1}& \tau_{r,1}^{a_2b_1} & 0&0\\
    \tau_{r,1}^{a_1b_2} & \tau_{r,1}^{a_2b_2} & 0&0
    \end{pmatrix},
    \end{aligned}
\end{equation}
whose entries are given by 
\begin{equation}
   \begin{aligned}
    \tau_{r,0}^{a_{1,2}b_{1,2}}&=t_{[3,3]},\quad 
    \tau_{r,1}^{a_1b_1}=\tau_{r,1}^{a_2b_2}=\left(1+e^{-ik_x}\right)t_{[3,3]},\\
    \tau_{r,1}^{a_{1,2}b_{2,1}}&=\left(e^{\pm i\theta}+e^{-ik_x}e^{\mp i\theta}\right)t_{[3,3]},
    \end{aligned} 
\end{equation}
where $t_{[3,3]}=\frac{2t_3}{11}$. Very much like in the $[2,3]$ case, we can map the minimal model of the zigzag ribbon unit cell into a  real space lattice, in this case a Creutz ladder, with $e^{\pm i\theta}$ Peierls factors in the intercell hopping terms amounting to an effective magnetic flux in alternating plaquettes [see Fig.~\ref{fig:4}(c)]. 
We perform a basis rotation using bonding/anti-bonding linear combination between the two pairs of zero modes localized in the same sublattice. This will break $C_3$ symmetry between the eigenfunctions in the new basis set $\{\ket{a_+},\ket{a_-},\ket{b_+},\ket{b_-}\}$ and the finite $\tau_{r,\sigma}^{ab}$ matrices now follow

\begin{equation}
   \begin{aligned}
    \tau_{r,0}^{a^+ b^+}&=2t_{[3,3]};\quad \tau_{r,1}^{a^\pm b^\mp}=\mp\left(1-e^{-ik_x}\right)\sqrt{3}it_{[3]}/2;\\
    \tau_{r,1}^{a^-b^-}&=3\tau_{r,1}^{a^+b^+}=\left(1+e^{-ik_x}\right)t_{[3,3]}/2.
    \end{aligned}
\end{equation}
In this new basis, the number of intracell hopping terms is reduced to a single $\tau_{r,0}^{a^+ b^+}$ and the ribbon unit cell effectively maps into a SSH chain composed of sites belonging to bonding states $\ket{a(b)_+}$ [see Fig.~\ref{fig:4}(d)]. Additionally, sites of anti-bonding states $\ket{a(b)_-}$ are connected to the main chain through closed loops, whose respective Peierls phase factors sum up to an effective zero transverse magnetic flux.

At this stage, the presence of a flat-band and the emergence of finite-energy edge states at a band touching point can be easily understood. When $k_x=0$, anti-bonding states become fully dimerized and contribute with $N_{rt}$ energies $\epsilon(k_x=0)=\pm3t_{[3,3]}$ to the flat-band. Meanwhile, the bonding states participating in the open SSH chain with hopping terms $\tau_{r,0}^{a^+ b^+}$ and  $\tau_{r,1}^{a^+ b^+}$ follow the generic form of the energy dispersion 
\begin{multline}
    \epsilon(k_x=0,k_n)=\\
    =\pm\sqrt{(\tau_{r,0}^{a^+ b^+})^2+(\tau_{r,1}^{a^+ b^+})^2+2\tau_{r,0}^{a^+ b^+}\tau_{r,1}^{a^+ b^+}\cos{(k_n)}}\\=t_{[3,3]}\sqrt{5+4\cos(k_n)}. 
\end{multline}
There are $N_{rt}-1$ complete dimer unit cells in the chain so that $k_n=\frac{\pi n}{N_{rt}}$ with $n=1,\dots,N_{rt}-1$. In this limit, $\tau_{r,0}^{a^+ b^+}>\tau_{r,1}^{a^+ b^+}$ and the open chain is in the topological regime, contributing to the missing chiral pair that completes the set of eigenstates for $k_x=0$. This pair localizes in the edge sites mapped as the bonding combination state of the two $C_3$ symmetric zero modes at the top and bottom edge $[3]$-triangulenes. When $k_x=\pm2\pi/3$, the hopping terms $|\tau_{r,1}^{a^- b^-}|=|\tau_{r,1}^{a^\pm b^\mp}|$ and the Peierls factors in the closed loops sum up to $\pi$, resulting in a topological transition where the chiral edge state pair pick up a finite opposite dispersion. At the zone boundary, $\tau_{r,1}^{a^+ b^+}$ vanishes and all bulk states are localized in two triangulene dimers contributing with energies $\epsilon(k_x=\pi)=\{\pm3t_{[3,3]},\pm t_{[3,3]}\}$ to the flat and dispersive bands, respectively. There are $N_{rt}-1$ of such 4-site clusters which effectively removes from each band one state of the accounted at $k_x=0$ [see Fig.~\ref{fig:4}(e)]. The missing four states stem from the additional dimerization at both boundaries responsible for the twice degenerate finite-energy edge states with $\epsilon_{\text{edge}}(k_x=\pi)=\pm\sqrt{3}t_{[3,3]}$. 

We recover the analytical form of the finite-energy edge states assuming solutions of the form $\ket{\epsilon_\text{edge}}=e^{ik_x}z_y^j\ket{u(z)}$ in the rotated basis set $\{\ket{a_+},\ket{a_-},\ket{b_+},\ket{b_-}\}$ for the following bulk Hamiltonian $h(k_x,z_y)$ of a $[3,3]$-triangulene crystal
\begin{equation}
    h(k_x,z_y)=
    t_{[3,3]}\begin{pmatrix}
    0& 0&\tau_{a_+b_+} &\tau_{a_+b_-} \\
    0& 0&\tau_{a_-b_+} &\tau_{a_-b_-} \\
    \tau_{a_+b_+}'& \tau_{a_-b_+}' & 0&0\\
    \tau_{a_+b_-}' & \tau_{a_-b_-}'& 0 & 0
    \end{pmatrix}, \label{h23edge}
\end{equation}
with entries
\begin{equation}
   \begin{aligned}
\tau_{a_-b_-}&=2+\cos(R_{1_x}k_x)z_y=3\tau_{a_+b_+}-2,\\
\tau_{a_\pm b_\mp}&=\pm\sqrt{3}\sin(R_{1_x}k_x)z_y,\\
\tau_{a_-b_-}'&=2+\cos(R_{1_x}k_x)z_y^{-1}=3\tau_{a_+b_+}'-2,\\
\tau_{a_\pm b_\mp}'&=\pm\sqrt{3}\sin(R_{1_x}k_x)z_y^{-1}.
    \end{aligned}
\end{equation}
The allowed $\{z_{y_n}\}$ solutions are obtained by requiring that the eigensolutions of the Hamiltonian in Eq.~\ref{h23edge} also satisfy the open boundary equations. In the rotated mapping version of Fig.~\ref{fig:4}(d), one can easily identify that the open conditions impose that the wave function vanishes at $b_+$ sites for the 0th dimer and at $a_+$ sites for the $(N_{rt}+1)$ dimer, that is $\braket{a/b_+|\epsilon_\text{edge}}=0$. This leads to two admissible solutions $z_y=\{\cos(R_{1_x}k_x),\sec(R_{1_x}k_x)\}$ that recover the decaying behavior from the top and bottom edge. The analytical form of the energy dispersion for the finite-energy edge states in $[3,3]$-triangulene ribbons follows
\begin{equation}
    \epsilon_\text{edge}=\pm \sqrt{3}t_{[3,3]}\sqrt{2-\cos(2R_{1_x}k_x)}.
\end{equation}
All analytical results provided in the former discussion are in agreement with the numerical calculations of zigzag ribbons considering the reduced model but not with the complete model. In the latter, we notice an overall broadening of the narrow bands around zero-energy and a small dispersion of the flat-bands in the vicinity of $\Gamma$, implying a small influence of higher/lower energy bands. 

\subsection{Zak phase analysis}

For $[3,3]$-triangulene zigzag ribbons, the loop around the BZ in the direction perpendicular to the edge has at most four singularity points of degeneracy: two between the dispersive bands at $\{K,K'\}$ and two between the dispersive and flat-bands at the $\Gamma$-point. These degeneracy points were disregarded in the calculation of the Zak phase for each band (see methods in appendix \ref{SMB}). The pair of zero-energy states in the midgap is predicted by the sum of the phases for the two lowest bands $\mathcal{Z}(k_x)=\mathcal{Z}_1+\mathcal{Z}_2$. Results show a topological non-trivial phase for $k_x \in [-2\pi/3,2\pi/3]$ and a trivial phase otherwise.  

The topology of the lowest and highest bandgaps were also evaluated. In the former, the total phase is given by the Zak phase of the lowest band $\mathcal{Z}=\mathcal{Z}_1$ and in the latter one should account for the three bands below the gap $\mathcal{Z}=\sum_{j=1}^3\mathcal{Z}_j$. Both phases are found to be always $\pi$-quantized, thus predicting the two pairs of dispersive edge states.

\section{$[4,4]$-triangulene ribbons}

In this section we introduce the results for the energy dispersion of $[4,4]$-triangulene ribbons using the tight-binding complete model and the effective reduced model on the basis of the zero modes. Unlike the three cases discussed so far, $[4,4]$-triangulene crystals have a gap at $E=0$. We focus our analysis on the localization of the edge modes found in both configurations and on the topological properties of the bandgaps in the band structure of zigzag ribbons, whose characterization can be equally reproduced in armchair ribbons.  

 \begin{figure*}[t]
     \centering
     \includegraphics[width=1\textwidth]{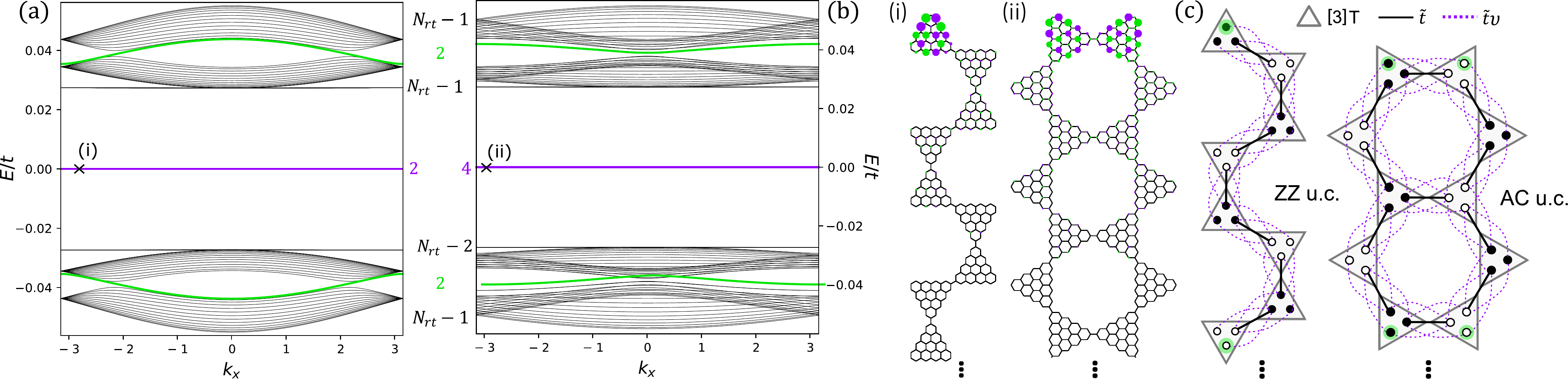}
     \caption{Model for $[4,4]$-triangulene ribbons with $N_{rt}=20$ dimers in the unit cell. The numerical results of the energy dispersion for the (a) zigzag and (b) armchair configurations show the presence of two types of midgap edge states, the ones highlighted in purple [green] have dispersionless [dispersive] zero [finite] edge states and localize in (i) a single zigzag edge triangulene or (ii) armchair edge dimer with small weights in adjacent triangulenes. The ribbon unit cell effectively maps into a (c) 2-orbital kagomé ribbon in the corner localized basis. Green highlight shows edge localization for $\upsilon=0$.}
     \label{fig:5}
 \end{figure*}

\subsection{Complete model}
The energy bands of $[4,4]$-triangulene ribbons in Fig.~\ref{fig:5}(a,b) feature two copies of the Kagome ribbon bulk bands separated by an energy gap at half-filling. The group of flat-bands, whose states localize in rings composed of six $[4]$-triangulenes, is degenerate at the $\Gamma$-point with a group of dispersive graphene-like bands. At the Dirac points between the latter groups, twice degenerate dispersive edge states emerge and are maximally localized at $k=\pi$ in the outermost edge dimers for zigzag boundaries, surviving for $|k|\in [2\pi/3,\pi]$. Armchair finite-energy edge states have a larger decay through the bulk even at the zone boundary.

Contrary to the $[3,3]$ case, the two identified band touching points between the $2(N_{rt}-1)$ [$2(N_{rt}-2)$] flat-band groups and the $2(N_{rt}-1)$ dispersive band groups at the $\Gamma$-point in the zigzag (armchair) configuration do not generate finite-energy dispersive edge states. The missing $2$ ($4$) states are found in the gap between the electron-hole symmetric bands, with
no dispersion, both for zigzag and armchair ribbons. In the zigzag case, these edge states have $E=0$, whereas in the armchair case this non-dispersive band features a very small width-independent energy splitting, likely due to intraedge hybdirization. As can be seen in Fig.~\ref{fig:5}(i,ii), their $k$-independent wave function is compactly localized in the majority sublattice at the top and bottom zigzag edge $[4]$-triangulene and in the armchair edge dimer with small weight in adjacent $[4]$-triangulenes. 

\subsection{Reduced model}

The degeneracy of the zero modes in $[n]$-triangulenes for $n>2$ allow one to choose a suitable representation for its set. In the particular case of $[4]$-triangulenes accommodating $n_{a/b}-1=3$ zero modes, these can be represented as the eigenstates of the 3-fold symmetry operator of Fig.\ref{fig:Fig1}(b) or rearranged into three orbitals related by a $2\pi/3$ rotation  \cite{Tamaki2020}. We use the analogue of a tight-binding ring of three sites where each site represents a state that is essentially populating one corner of the $[4]$-triangulene. The new corner localized basis $\ket{c_j}$ can be constructed from the $C_3$ symmetric zero modes $\ket{\omega}$ with $\omega=\{0,\pm2\pi/3\}$ from the linear combination $\ket{\omega}=\frac{1}{\sqrt{3}}\sum_{j=1}^{3}e^{i\omega j}\ket{c_j}$. 
The finite elements of the effective zigzag ribbon Hamiltonian in this basis are given by the intra and intercell contributions
\begin{equation}
    \tau_{r,0}=\begin{pmatrix}
        0^{[n_a-1]}&\tau_{r,0}^{c}\\
        \tau_{r,0}^{c}{}^\dagger&0^{[n_b-1]}
    \end{pmatrix},\quad
    \tau_{r,1}=\begin{pmatrix}
        0^{[n_a-1]}&0\\
        \tau_{r,1}^{c}{}^\dagger&0^{[n_b-1]}
    \end{pmatrix},
\end{equation}
whose entries follow 
\begin{equation}
\begin{aligned}
   \tau_{r,0}^{c}&=t_{[4,4]}\begin{pmatrix}
    1&\upsilon&\upsilon\\ 
    \upsilon&0&0\\  
    \upsilon&0&0
   \end{pmatrix},\\
   \tau_{r,1}^{c}&=t_{[4,4]}\begin{pmatrix}
    0&\upsilon&\upsilon e^{-ik_x}\\ 
    \upsilon&1&\upsilon(1+e^{-ik_x})\\  
    \upsilon e^{-ik_x}&\upsilon(1+e^{-ik_x})&e^{-ik_x}
   \end{pmatrix},
\end{aligned}
\end{equation}
where $t_{[4,4]}=0.35t_3$ and $\upsilon t_{[4,4]}=0.033t_3$ describe first and second neighboring overlap between adjacent corner states, respectively. The non-vanishing second neighbor term is a byproduct of the non-zero weights of the corner localized states at sites close to the other two corners. Next order terms are  smaller than $\upsilon t_{[4,4]}$ and can be discarded.

The zigzag unit cell of the reduced model is depicted in Fig.~\ref{fig:5}(c). For $\upsilon=0$, every corner localized state only hybridizes with a single corner state on the adjacent triangulene. In this dimerized limit, a dispersionless valence bond solid is formed from the corresponding bonding and antibonding states. These states are centered at the midpoint between neighboring corners and are weakly coupled by a non-zero $\upsilon t_{[4,4]}$, effectively tailoring kagome ribbon lattices. Indeed, the low-energy spectrum gives a pair of kagome subbands from these bonding and antibonding states. 

As expected, the effective hopping terms do not connect corner states within each [4]-triangulene nor belonging to the same sublattice. Therefore, each triangle [see  Fig.~\ref{fig:5}(c)] belongs to a different sublattice, preserving the chiral symmetry of the model. Zigzag edge states obey this symmetry as they remain zero-energy states with increasing $\upsilon$. In the armchair configuration and when $\upsilon = 0$, two edge states with support in both sublattices emerge on the same edge. A non-zero $\upsilon$ term couples these states, leading to energy splitting of order $\upsilon^2$. Higher order terms make the edge state band dispersive with small bandwidth at least of the order of $\upsilon^4$. 
The limit of $\upsilon\gg 1$ does not lead to dimerization and therefore there is no gap reopening. For each $k_x$ in the ribbon, $\upsilon=0$ implies a dimerized chain similar to the $t_1=0$ limit of the SSH chain.

\subsection{Zak phase analysis}
We ought to evaluate the topology of both the insulating midgap and the two bandgaps between the Dirac points of the $[4,4]$ zigzag ribbon band structure. We identify six singularity points of degeneracy: four between the dispersive bands at $\{K,K'\}$ and two between the dispersive and flat-bands at the $\Gamma$-point. The overlap terms at these points were dropped from the numerical calculation of the Zak phase for the corresponding bands. 

The pair of zero-energy edge states in the midgap is predicted by the sum of the phases for the three lowest bands $\mathcal{Z}=\sum_{j=1}^3\mathcal{Z}_j$. Results show a topological non-trivial phase for all $k_x$. 

The topology of the lowest and highest bandgaps were also evaluated, having into account the presence of degeneracies at the Dirac cones. For the lowest bandgap, the total phase is given by the Zak phase of the lowest band $\mathcal{Z}=\mathcal{Z}_1$ and for the highest bandgap one should account for the five bands below the gap $\mathcal{Z}=\sum_{j=1}^5\mathcal{Z}_j$. Both phases are found to be $\pi$ for $|k_x|\in [2\pi/3,\pi]$ and trivial when $k_x\in [-2\pi/3,2\pi/3]$, thus effectively predicting the two pairs of dispersive edge states. 

\section{Conclusions}

We have carried out an extensive study on the properties of edge states on a class of models that generalize the conventional honeycomb tight-binding Hamiltonian to the case where each site has several degenerate modes, eigenstates of the $C_3$ symmetry operator. This model is relevant to describe many systems, including triangulene crystals, the main motivation for our work. We have discussed in detail four classes of triangulene crystals by examining the band structure of one-dimensional triangulene ribbons. The construction of simpler one-dimensional effective models with $k$-dependent hopping parameters, such as the diamond chain for the $[2,3]$-triangulene and the creutz ladder for the $[3,3]$-triangulene, allowed one to understand the regimes where these edge states emerge. 
We have found exotic edge phenomena, absent in the conventional honeycomb tight-binding model of graphene, deriving from the non-trivial features of the two-dimensional bands of triangulenes: 
\begin{itemize}
    \item[-] A pair of edge states living in the same $[2]$-triangulene edge of the $[2,3]$ zigzag ribbon: a consequence of the zero mode mismatch between triangulenes results in edge dimerization for the top edge and an additional zero-energy edge state in the bottom edge at the zone boundary.
    \item[-] Edge states occurring at flat-parabolic touching-points and bearded edge states when degenerate with a bulk state from the Dirac point of $[3,3]$ ribbons, resulting from edge dimerization on both boundaries and the non-trivial regime of an SSH chain, respectively. 
    \item[-] Zero-energy edge states in the bandgap at half-filling of the $[4,4]$ bands, that arise as a consequence of unpaired zero-modes in both edges. 
    \item[-] Edge states surviving in armchair ribbons whenever the dimensions of the triangulene dimer $n_{a}+n_b>4$.
\end{itemize}

We have assessed the topological origin of zigzag edge states using the single-band Zak phase formalism. Although the bulk-boundary correspondence for finite-energy dispersive edge states remains an open question, we argue that the presence of all edge states in $[3,3]$ and $[4,4]$ zigzag ribbons was successfully linked to non-trivial Zak phase values, including the dispersive finite-energy edge states appearing between degenerate bands. The same approach can be applied to armchair edge states for the right choice of unit cell and $k_\parallel$-direction. 

Our results reveal the prominence of peculiar properties in $C_3$ symmetric multi-orbital honeycomb lattices. Full concurrence between the complete and reduced tight-binding versions correlates the appearance of edge states to the low-energy bands associated with the zero modes of single triangulenes. We argue that the agreement between the complete and the mininmal model stresses the crucial role played by the interplay between the $C_3$ symmetry of the high-fold fermions in each site of the honeycomb lattice and the $C_3$ symmetry of the lattice. Therefore, we expect that our analys will be relevant for other multi-orbital models with $C_n$-symmetry on other crystalline arrangements.  \\

\acknowledgements

L.M. acknowledges the financial support from FCT (grant No. PTDC/FIS-MAC/2085/2020).
L.M. and J.F.-R. acknowledge support from the European Union (Grant FUNLAYERS-101079184).
We acknowledge fruitful discussions with João Henriques and António Costa. 
J.F.-R. acknowledges financial support from Generalitat Valenciana (Prometeo2021/017 and MFA/2022/045), from SNF Sinergia (Grant Pimag, CRSII5-205987) and MICIN-Spain (Grants PID2019-109539GB-C41 and PRTR-C1y.I1). L.M. and R.G.D. developed their work within the scope of Portuguese Institute for Nanostructures, Nanomodelling and Nanofabrication (i3N) Projects No. UIDB/50025/2020, No. UIDP/50025/2020, and No. LA/P/0037/2020, financed by national funds through the Fundação para a Ciência e Tecnologia (FCT) and the Ministério da Educação e Ciência (MEC) of Portugal. 

\appendix

\section{Edge states in armchair terminated ribbons}\label{SMA}

\begin{figure*}
    \includegraphics[width=1\textwidth]{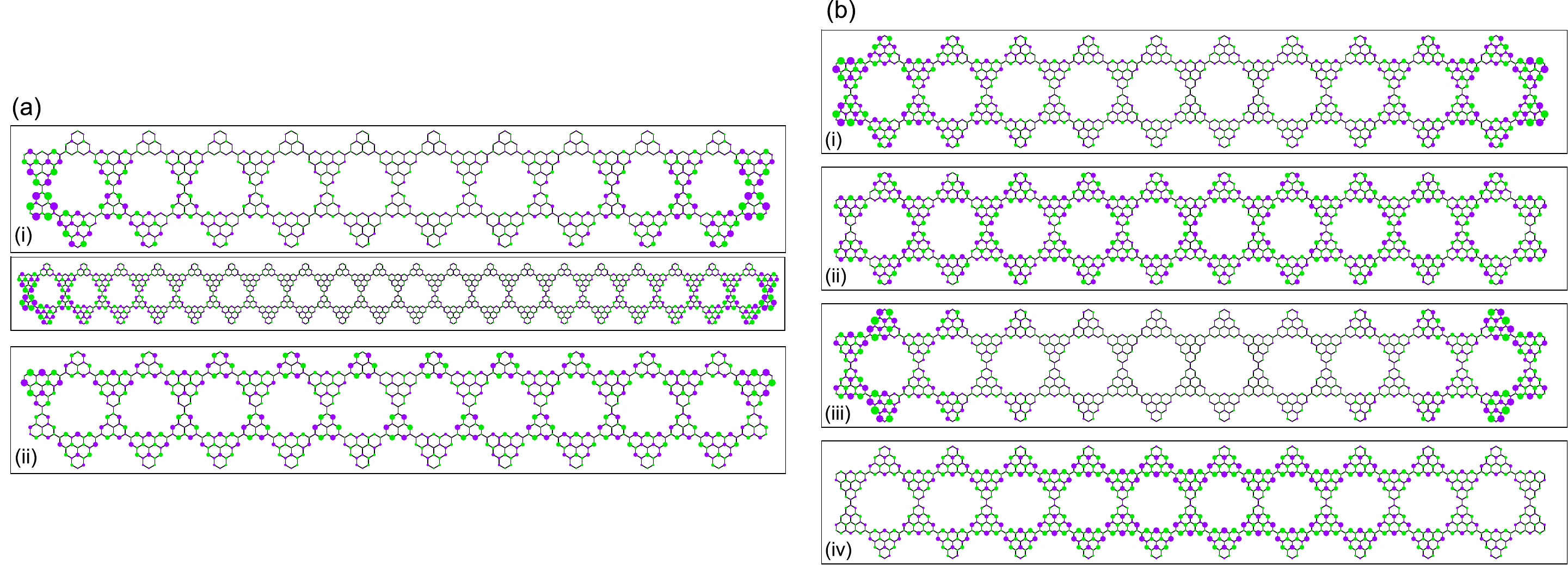}
    \caption{Profile of the wavefunction for the dispersive edge states found in (a) $[2,3]$ armchair ribbons at (i) $k_y=\pi$ and (ii) at $k_y=0$; (b) $[3,3]$ armchair ribbons at (i), (iii) $k_y=\pi$ and at (ii), (iv) $k_y=0$. We show the probability distribution of the lowest energy edge state in the unit cell of the armchair ribbon periodic in the $y$-direction with $N_{rt}=21$ and $N_{rt}=41$ dimers for a(i).}
    \label{fig:ApAB}
\end{figure*}

The observed valley mixing on armchair graphene largely influences the $\pi$-electron states near the Fermi level and accounts for the absence of edge states  \cite{Wakabayashi2009}. Herein, the isospin of the fermions within each individual triangulene for $n_{a/b}>2$ recovers edge localization, despite featuring no intra-edge sublattice imbalance (edges terminate with A-B triangulene dimers). In the following appendix, we briefly discuss the identified midgap edge states in $[2,3]$ and $[3,3]$ triangulene armchair ribbons. 

In Fig.~\ref{fig:ApAB}(a) we represent the lower energy state of the electron-hole pair of dispersive edges states found in the gap between the dispersive and the flat subbands of the [2,3] armchair ribbon. Since inter-edge sublattice imbalance is absent in the armchair configuration, the edge state can be found in both sides of the ribbon and accounts for the double electron-hole edge state pair at the zone boundary. As one can see in Fig.~\ref{fig:ApAB}a(i) for an armchair ribbon with $N_{rt}=21$ and a larger ribbon with $N_{rt}=41$ dimers, this state is mostly localized in the outermost $[2]$-triangulene and two adjacent $[3]$-triangulenes. Similar to its zigzag counterpart, the localization length of this state increases as one approaches the $\Gamma$-point [see Fig.~\ref{fig:ApAB}a(ii)].

The probability distribution of the edge states found in the three gaps of $[3,3]$ armchair ribbons is illustrated in Fig.~\ref{fig:ApAB}(b). We find a total of two doubly degenerate electron-hole pairs of dispersive edge states. The first pair is localized in the midgap of the bulk dispersive subbands while the second pair can be found in the gap between the dispersive and flat subbands. These modes are present for all momenta, are maximally localized at the zone boundary in both sides of the ribbon [see Fig.~\ref{fig:ApAB}b(i,iii)] and the exponential decay through the bulk increases closer to the $\Gamma$-point [see Fig.~\ref{fig:ApAB}b(ii,vi)]. The dispersive edge states in the central gap populate the outermost dimers while the lowest energy pair localizes in the second to the last row of dimers, with a gradual decrease towards the bulk.

\section{Numerical calculation of the total Zak phase for [3,3] and [4,4] ribbons} \label{SMB}

\begin{figure*}
    \centering
    \includegraphics[width=0.85\textwidth]{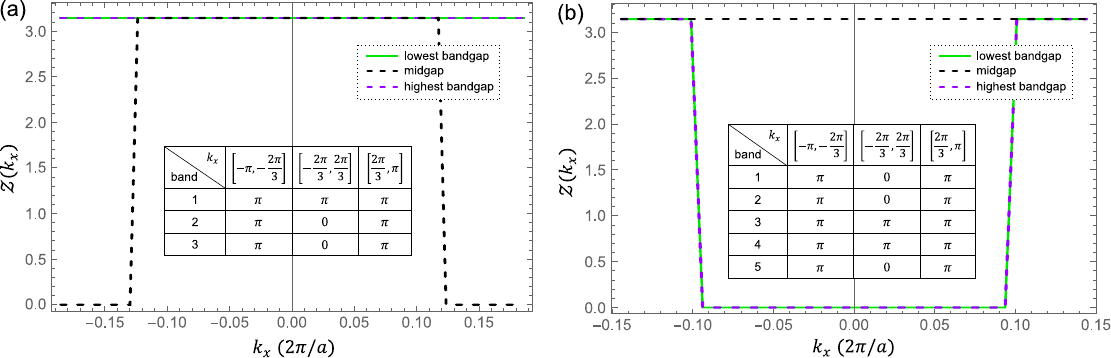}
    \caption{Numerical calculation of the total Zak phase for the identified bandgaps on the band structure of (a) $[3,3]$ and (b) $[4,4]$ triangulene ribbons. Inset tables display the $\mathcal{Z}_n$ for each band numbered in increasing energy order for the relevant intervals of momentum $k_x$. The phase of the highest band is not shown here as is irrelevant for the calculation of $\mathcal{Z}$.}
    \label{fig:ApC}
\end{figure*}

We have used the single band formalism to predict the topological phases of all bandgaps of $[3,3]$ and $[4,4]$ triangulene ribbons. This often poses some challenges in the presence of degenerate bands if no considerations are taken on the gauge choice. We have treated the presence of singular points of degeneracy for each $k_\parallel$ as follows: we make use of a smooth gauge choice where phases are roughly uniformly distributed throughout the loop. This distribution becomes less uniform close to topological transition points. The gauge transformation is possible by finding, for each band, the non-zero component of $\ket{u(k)}$ that has the smallest change throughout the $k$-scanning in the Wilson loop and choosing the gauge where this component is real. 
The advantage of using a smooth gauge is that it allows for the calculation of the Zak phase when there is finite number of band crossings. These points can be dropped out numerically, but one must take into account the change in the ordering of the bands at these crossing points using for example a largest overlap criterium to correctly order the bands. For both the [3,3] and [4,4] ribbons, the bands considered for the estimation of the total Zak phase have at most three singularity points of degeneracy: two between the dispersive bands at $k_\bot=2\pi/3$ and one between dispersive and a flat-band at $k_\bot=0$. Disregarding the overlaps elements in these points will not drastically affect the calculation of the total phase. 

We also have to note that the value of Zak phase of each band may have a difference of $\pm \pi$ for different choices of the bulk unit cell. We make a choice of unit cell dimer such that it allows the full reconstruction of the lattice including its specific zigzag edge.

The total Zak phases for the three identified bandgaps of the $[3,3]$ and $[4,4]$ ribbons are depicted in Fig.~\ref{fig:ApC}(a,b), respectively. These were determined by the sum of the Zak phases for each band below the corresponding bandgap, whose values can be found in the inset tables in Fig.~\ref{fig:ApC}. Specifically for the lowest bandgap, the total Zak phase is equivalent to $\mathcal{Z}_1$ of the lowest energy band (which is the only band found below this gap).  
We can see that the non central bandgaps of [3,3] ribbons are always $\pi$ while the midgap takes a non-trivial topological phase when $k_x \in [-2\pi/3,2\pi/3]$, agreeing with the localization of the states found in these gaps. On the contrary, only the midgap of $[4,4]$ ribbons is $\mathcal{Z}=\pi$ independent of $k_x$. The remaining gaps have non-trivial topology in the complementary interval $|k_x|\in [2\pi/3,\pi]$, which effectively predicts the pair of dispersive edge states.

Singular points of degeneracy are present when $k_x=\{0,\pm2\pi/3\}$. In [3,3] ribbons, $\mathcal{Z}_n(0)$ possesses two band crossings at $k_y=0$ between bands $1,2$ and $3,4$ and $\mathcal{Z}_n(\pm2\pi/3)$ with $n=2,3$ has two degenerate points at $k_y=\pm \pi$. In the case of [4,4] ribbons, we find two degenerate points between bands $2,3$ and $4,5$ that will affect the overlap elements for $k_y=0$ and we also find four band crossings for $k_x=\pm2\pi/3$ between the two highest and lowest bands at $k_y=\pm \pi$. 
We have disregarded the overlap elements of all these degeneracy points which causes the total phase $\mathcal{Z}(0,\pm 2\pi/3)$ to have a minimal deviation from the  $0\,\wedge\,\pi\mod2\pi$ quantization expected for inversion symmetric models. This error gets mitigated if we consider a smaller $\Delta k_y$.  

\bibliographystyle{apsrev4-2}
\bibliography{bibliography}

\end{document}